\documentstyle[twocolumn,floats,epsfig,ifthen,aps,prb]{revtex}

\begin{document}
\draft
\title{Shot noise at hopping via two sites} 
\author{Yusuf A. Kinkhabwala and Alexander N. Korotkov} 
\address{Department of Physics and Astronomy, 
State University of New York, Stony Brook, New York 11794-3800 }

\date{\today}

\maketitle

\begin{abstract}
        The average current and the shot noise at correlated sequential 
tunneling via two localized sites are studied. At zero temperature 
the Fano factor averaged over the positions and energies of sites 
is shown to be 0.707. 
The noise dependence on temperature and frequency is analyzed numerically. 
\end{abstract}



\narrowtext

        Shot noise in mesoscopic structures has been the subject of 
thorough studies in the recent past.\cite{Kogan,deJong,Blanter} 
In particular, a theory of shot noise at tunneling 
of single electrons correlated due to Coulomb blockade effects 
\cite{Av-Likh} has been well developed
\cite{Kor-92,Kor-94,Hershfield,Galperin,Kor-98,Matsuoka}
and verified experimentally.\cite{Birk} 
        Recently the first attempts have been made 
\cite{Kor-Likh,Sverdlov} to extend 
this theory to hopping transport \cite{Shklovskii} which 
can be formally considered as a special case of correlated single-electron
tunneling. 
        
In a typical hopping situation there is a considerable 
$1/f$ contribution at low frequencies (see Ref.\ \cite{Kogan-hop} 
and references therein), so one can discuss the shot noise only 
at sufficiently high frequencies. The $1/f$ noise at hopping is mainly 
due to electron-electron interaction: the slowly evolving trapped charge 
configurations can significantly affect the current 
through nearby channels. The $1/f$ component is absent at hopping 
through noninteracting 1D chains of sites \cite{Kor-Likh} (while the slow 
fluctuations of the chain parameters due to external traps can restore 
this component).
In the present paper we consider hopping through very short chains 
which are just pairs of sites,  
and assume that the parameters of these pairs do not fluctuate in time, 
so that the noise does not have $1/f$ contribution. 

        For two-site hopping we basically follow the model introduced
by Glazman and Matveev.\cite{Glazman} The only difference is that 
we take into account the correlation between tunneling events neglected
in Ref.\ \cite{Glazman} (in this respect our model is closer to the 
model of Ref.\ \cite{Raikh}). 
Using the methods developed in Ref.\ \cite{Kor-94}
we calculate the current $I$ and the current spectral density 
$S_I(\omega )$ for an individual two-site channel. 
The summation over many parallel channels with random parameters 
is done similar to Ref.\ \cite{Glazman}. 
The main object of our study is the Fano factor $F$ (the low frequency noise 
normalized by the Schottky value $S_I=2eI$).    
We will show that at zero temperature 
the Fano factors for individual two-site channels range from $5/14$
to $1$, while after averaging we get ${\overline F}=0.707$ (the similar 
problem for one-site channels has been considered in Ref.\ \cite{Nazarov} 
with the result ${\overline F}=3/4$). For a finite temperature $T$ the Fano 
factor can be calculated numerically; after the averaging we obtain  
$\overline F$ as a function of the ratio $T/eV$ where $V$ is the voltage
between electrodes. 

\begin{figure} 
\begin{center}
\epsfig{figure=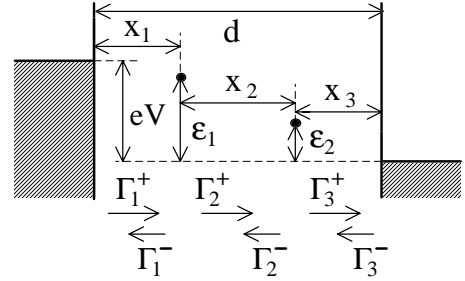,scale=0.3}
\vskip-0.2cm
\end{center}
\caption{ Schematic of two-site tunneling channel. 
 }
\label{schematic}\end{figure}

        The schematic of a two-site channel is shown in Fig.\ \ref{schematic}.
The thickness $d$ of an insulating layer between two metallic electrodes
is assumed to be much greater than the electron localization radius $a$, 
and we use the model of sequential (incoherent) hops of single electrons.   
The lengths of the left and right hops are $x_1$ and $x_3$, respectively, 
while the hop between two sites has the length $r_2=(x_2^2+y^2)^{1/2}$ where
$x_2=d-x_1-x_3$ and $y$ is the shift of site positions in the plane parallel 
to electrodes. Each site can be occupied by at most one electron and 
the effect of electron spin is neglected (the case of double-degeneracy 
due to spin will be discussed later). The tunneling rates from electrodes 
to empty sites (tunneling to nearest neighbor only, see Fig.\ 
\ref{schematic}) are assumed to be 
        \begin{eqnarray} 
&& \Gamma_1^+ = \Gamma_0 \exp(-2x_1/a) f(-eV+\varepsilon_1) , 
        \nonumber\\
&& \Gamma_3^- = \Gamma_0 \exp(-2x_3/a) f(\varepsilon_2),  
        \label{Gamma_in}\end{eqnarray} 
where superscripts indicate the direction of tunneling, 
$\varepsilon_{1}$ and  $\varepsilon_{2}$ 
are the site energies counted from the Fermi level of the right electrode, 
and $f(\varepsilon )=[1+\exp(\varepsilon /T)]^{-1}$ is the Fermi function.
Similarly, the rates of tunneling from occupied sites to neighboring 
electrodes
are 
        \begin{eqnarray}
&& \Gamma_1^- = \Gamma_0 \exp(-2x_1/a) f(eV-\varepsilon_1) ,
        \nonumber\\
&& \Gamma_3^+ = \Gamma_0 \exp(-2x_3/a) f(-\varepsilon_2). 
        \label{Gamma_out}\end{eqnarray} 
Notice that we have neglected the Coulomb interaction of electrons 
on different sites (energies $\varepsilon_{1,2}$ do not depend on
the occupation of neighboring site).  
The rate of inelastic tunneling between the sites depends on the energy 
difference $\Delta \varepsilon =\varepsilon_1 -\varepsilon_2$, and for 
$|\Delta \varepsilon|$ much smaller than $\hbar s/a$ (where $s$ is
the sound velocity) can be calculated 
as \cite{Shklovskii} 
        \begin{equation}
\Gamma_2^\pm = \alpha\Gamma_0 \exp (-2r_2/a) \, \frac{\pm \Delta \varepsilon} 
{1-\exp(\mp \Delta\varepsilon /T)},  
        \label{Gamma_2}\end{equation}
where the dimensional factor $\alpha$ describes the relative strength of 
phonon-assisted tunneling compared to ``resonant'' tunneling assumed 
in Eqs.\ (\ref{Gamma_in})--(\ref{Gamma_out}). 

\begin{figure} 
\begin{center}
\vskip-0.2cm
\hspace{-0.3cm}\epsfig{figure=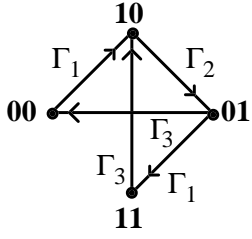,scale=0.17}
\vskip-0.2cm
\end{center}
\caption{ Graphical representation of the master equation for zero 
temperature.} 
\label{master-eq}\end{figure}

        Let us start with zero temperature case. Then the transport is
possible only if $eV > \varepsilon_1 > \varepsilon_2 >0$, and electrons 
move only in one direction, $\Gamma_1^-=\Gamma_2^-=\Gamma_3^-=0$
(for simplicity we omit the superscript ``$+$'', $\Gamma_i \equiv 
\Gamma_i^+$). The 
kinetic  (``master'')  equation  in  this  case  can   be   represented 
graphically
by Fig.\ \ref{master-eq}.  The configuration space consists of four 
charge states of the two-site system which 
are denoted as 00, 01, 10, and 11, while arrows show transitions between 
them. The graphical representation of the master equation in a  
relatively small configuration space is a very convenient tool and often
allows straightforward calculation of the average current and zero-frequency
spectral density (see, e.g., Ref.\ \cite{Kor-set-an}).

        The basic idea of the method \cite{Kor-94} is to consider the 
random ``travel'' of the system state within the configuration space and 
divide the duration of this stochastic process into blocks 
which start and end in a specific charge 
state. Because of the Markovian property of the process (absence of memory)
these blocks are mutually uncorrelated, so the averaging over the blocks 
is rather simple. In the case of Fig.\ \ref{master-eq} let us choose the 
charge state 01 as the block divider. Then there are two types of blocks: 
$01\rightarrow 00\rightarrow 10\rightarrow 01$ (type 1) and   
$01\rightarrow 11\rightarrow  10\rightarrow  01$  (type  2),  while 
the blocks are additionally characterized by 
the time spent in each charge state. 

        The average current can be calculated \cite{Kor-94} as 
        \begin{equation}
I=e{\overline k}/{\overline \tau}, 
        \label{current}\end{equation} 
 where ${\overline \tau}$ is the average block duration and 
${\overline k}$ is the average number of electrons transferred between
electrodes  per block (the averaging is taken over a large number of blocks). 
        To calculate these average magnitudes let us notice that the blocks
of type 1 and type 2 have probabilities
        \begin{equation}
p_1=\Gamma_3/(\Gamma_1 +\Gamma_3), \,\,\, 
p_2=\Gamma_1/(\Gamma_1 +\Gamma_3),    
        \label{p_12}\end{equation}
and the average durations $\overline{ \tau_{1}}$ and
$\overline{ \tau_{2}}$  of the blocks of 
each type can be calculated as
        \begin{eqnarray}
&& \overline{ \tau_1}=(\Gamma_1 +\Gamma_3)^{-1} +\Gamma_1^{-1}+
\Gamma_2^{-1},
        \nonumber \\
&& \overline{ \tau_2}=(\Gamma_1 +\Gamma_3)^{-1} +\Gamma_3^{-1}+
\Gamma_2^{-1} 
        \label{tau_12}\end{eqnarray}
(notice that the average waiting time $(\Gamma_1 +\Gamma_3)^{-1}$ 
of the first hop is equal for both types). Taking into account that each
block corresponds to the transfer of one electron, $k_1=k_2=\overline{ k}=1$,
and calculating the average block duration 
        \begin{equation}
{\overline \tau} = p_1 \overline{ \tau_1} +p_2 \overline{ \tau_2},  
        \label{tau}\end{equation}
we finally obtain the formula for the average current:
\cite{interaction} 
        \begin{equation}
I= e \left( \frac{1}{\Gamma_2} + \frac{1+\Gamma_1 /\Gamma_3
+\Gamma_3/\Gamma_1} {\Gamma_1 +\Gamma_3} \right)^{-1}.
        \label{cur}\end{equation}

        This equation differs from the result of Ref.\ \cite{Glazman}
because the correlation between the occupations of two sites was
neglected in Ref.\ \cite{Glazman}. The correct equation for the current 
which coincides with Eq.\ (\ref{cur}) was obtained later in Ref.\ 
\cite{Raikh}.

        The same method as above can be used for the calculation of  
the low-frequency limit $S_I(0)$ of the current spectral density 
which is given by the general equation\cite{Kor-94}
        \begin{equation}
S_I(0)=\left(2/ \overline \tau \right) \left( e^2 \overline{ k^2} 
+I^2 \overline{ \tau^2} -2eI \overline{ k\tau} \right)  
        \label{Sgen}\end{equation}
(averaging is again over blocks) which in our case at zero
temperature simplifies to 
        \begin{equation}        
S_I(0)=2eI \left[ (\overline{ \tau^2}/\overline{ \tau}^2) -1 \right] .  
        \end{equation}
So, besides Eqs.\ (\ref{current})--(\ref{tau}) we also need to calculate 
$\overline{\tau^2}$: 
        \begin{equation}
\overline{\tau^2}=p_1 \overline{\tau_1^2} +p_2 \overline{\tau_2^2},
        \end{equation}
where because of Poissonian statistics of each tunneling event we have
        \begin{eqnarray}
\overline{\tau_1^2}-\overline{\tau_1}^2 = 
(\Gamma_1 +\Gamma_3)^{-2} +\Gamma_1^{-2}+\Gamma_2^{-2},
        \nonumber \\
\overline{\tau_2^2}-\overline{\tau_2}^2 = 
(\Gamma_1 +\Gamma_3)^{-2} +\Gamma_3^{-2}+\Gamma_2^{-2}.
        \end{eqnarray}

Combining these equations we finally obtain
        \begin{eqnarray}
S_I(0)= && 2eI \left( \frac{1}{\Gamma_2^2} +
\frac{ (1+R+R^{-1})^2 -4 }
{(\Gamma_1 +\Gamma_3)^2} \right)
        \nonumber \\
&& \times \left( \frac{1}{\Gamma_2} + \frac{1+R+R^{-1}} 
{\Gamma_1 +\Gamma_3} \right)^{-2} ,
        \label{S_I}\end{eqnarray}
where $R\equiv \Gamma_1/\Gamma_3$. 
        Analyzing the Fano factor $F\equiv S_I(0)/2eI$ one can see that 
the uniform case, $\Gamma_1=\Gamma_2=\Gamma_3$, provides $F=9/25$ 
which is not the minimum possible value. The minimum Fano factor is 
achieved at $\Gamma_1=\Gamma_3=5/6\times \Gamma_2$ and equal to 
$F_{min}=5/14$ (it is still noticeably larger than the naive 
estimate $F=1/3$). 
The maximal value $F_{max}=1$ is obviously achieved when one of 
the rates $\Gamma_i$ is much smaller than two other rates. 

        Following Ref.\ \cite{Glazman} let us assume many two-site 
channels ``in parallel'' and find the total current $I_\Sigma$ and 
spectral density $S_{I,\Sigma}(0)$ integrating over channels with different 
site positions $x_1$, $x_3$, $y$ and  different energies $\varepsilon_1$ 
and $\varepsilon_2$. Assuming a  sufficiently thick insulating layer 
we may approximate the distance between sites as 
$r_2\simeq x_2 +y^2/2{\tilde x}_2$, where ${\tilde x}_2$ corresponds
to the channel with the maximum current. 
For such a channel $y=0$ and $\Gamma_1=\Gamma_2=\Gamma_3=
\Gamma_0 \exp(-2d/3a) (\alpha eV)^{1/3}$ [see Eq.\ 
(\ref{cur}) and Eqs.\ (\ref{Gamma_in})--(\ref{Gamma_2})] which gives 
$\tilde x_2=(d/3)+(a/3)\ln (\alpha eV) $. At zero temperature
the total current can be calculated as 
        \begin{eqnarray} 
I_{\Sigma} && = n^2 A \int_{-\infty}^\infty d\xi_1 
\int_{-\infty}^\infty d\xi_3 \int_0^\infty 2\pi y \,  dy 
        \nonumber \\ 
&& \times  \int_0^{eV} d\Delta\varepsilon \,\, (eV-\Delta\varepsilon ) \, 
 I(\xi_1 ,\xi_3 ,y, \Delta \varepsilon) ,
        \label{Isum}\end{eqnarray}
where $n$ is the density of states, $A$ 
is the area [$A\gg d^2$, $A\gg n^{-2} a^{-3}d^{-1}(eV)^{-2}$], 
the $x$-positions of the sites 
are measured from the optimal values, $\xi_i=x_i-\tilde x_i$,
$\tilde x_1=\tilde x_3=(d-\tilde x_2)/2$ (the integration is extended
to infinity since $d\gg a$), and 
the current $I$ is given by Eq.\ (\ref{cur}). 

	Using the relation $ I(\xi_1 ,\xi_3 ,y, \Delta \varepsilon) 
=\exp (- 2\delta /a)\, I(\xi_1 -\delta ,\xi_3-\delta ,0, \Delta \varepsilon)$
where $\delta =y^2/6{\tilde x}_2$, 
it is easy to show that the integration 
$\int_0^\infty 2\pi y \, dy $ gives the factor $3\pi a{\tilde x}_2$. 
Calculating the integral over $\Delta \varepsilon$ analytically 
and integrals over $\xi_1$ and $\xi_3$ numerically, we get the result 
        \begin{equation}
I_\Sigma = 5.237 \, e \Gamma_0 \, n^2 A a^3 {\tilde x}_2 \, 
        \exp (-2d/3a)\,  \alpha^{1/3} (eV)^{7/3}. 
       \label{Isum2} \end{equation}
Notice that the scaling $I_\Sigma \propto V^{7/3}$ is the same as
in the model which neglects correlations.\cite{Glazman} 

        A similar sum over different two-site channels can be calculated
for the current spectral density at zero frequency [just 
replacing $I$ in Eq.\ (\ref{Isum}) with $S_I(0)$ given by Eq.\
(\ref{S_I})]. Integrating analytically over $y$ and $\Delta \varepsilon$ 
and numerically 
over the two remaining variables we obtain the following average Fano factor 
at zero temperature: 
        \begin{equation}
\overline{F} \equiv S_{I,\Sigma}(0)/2eI_\Sigma = 0.7074. 
        \label{Faver}\end{equation}

        Let us now consider the finite-temperature case. Our method 
for calculation of $I$ and $S_I(0)$ based on the analysis of blocks 
can still be easily applied if $\Gamma_2^- \neq 0$ while 
$\Gamma_1^-=\Gamma_3^-=0$ (this situation occurs when  
$\varepsilon_1$ and $\varepsilon_2$ are well inside 
the energy strip defined by the Fermi levels of the electrodes). In this case 
the current $I$ and the spectral density $S_I(0)$ are given by Eqs.\ 
(\ref{current}) and (\ref{Sgen}), where\cite{e-e_2}
        \begin{eqnarray}
&& {\overline \tau} = p_1 \overline{ \tau_1} +p_2 \overline{ \tau_2}
+ p_3 \overline{ \tau_3}, \,\,\,\,\, 
{\overline k}=\overline{k^2}=p_1+p_2 , 
        \nonumber \\ 
&& p_1=\Gamma_3^+/\Gamma_\Sigma, \,\, p_2=\Gamma_1^+/\Gamma_\Sigma, \,\, 
p_3=\Gamma_2^-/\Gamma_\Sigma, 
        \nonumber \\
&& \overline{\tau_1} = 1/\Gamma_\Sigma + 1/\Gamma_1^+ 
        + 1/\Gamma_2^+, \,\,\, 
\overline{\tau_3} = 1/\Gamma_\Sigma + 1/\Gamma_2^+ ,
        \nonumber \\
&& \overline{\tau_2} = 1/\Gamma_\Sigma + 1/\Gamma_3^+ +1/\Gamma_2^+ , \,\,\,
 \Gamma_\Sigma = \Gamma_3^+ +\Gamma_1 ^+ +\Gamma_2^- ,
        \nonumber \\
&&\overline{\tau^2}=p_1 \overline{\tau_1^2} +p_2 \overline{\tau_2^2}
+p_3 \overline{\tau_3^2}, \,\,\,
\overline{ k\tau} = p_1\overline{\tau_1}+p_2\overline{\tau_2}, 
        \nonumber \\
&& \overline{\tau_1^2}-\overline{\tau_1}^2 = 
    (\Gamma_\Sigma )^{-2} +(\Gamma_1^+)^{-2}+(\Gamma_2^+)^{-2},
        \nonumber \\
&& \overline{\tau_2^2}-\overline{\tau_2}^2 = 
(\Gamma_\Sigma )^{-2} +(\Gamma_3^+)^{-2}+(\Gamma_2^+)^{-2}.
        \nonumber \\
&& \overline{\tau_3^2}-\overline{\tau_3}^2 = 
(\Gamma_\Sigma )^{-2} + (\Gamma_2^+)^{-2}. 
        \end{eqnarray}

\begin{figure} 
\begin{center} 
\vskip-0.2cm
\hspace{-0.3cm}\epsfig{figure=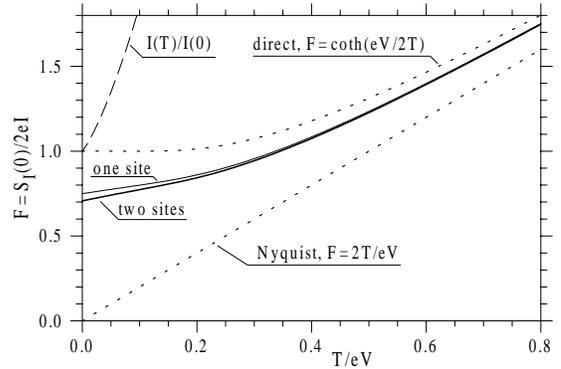,scale=0.4} 
\vskip-0.2cm
\end{center}
\caption{ The Fano factor ${\overline F}$ averaged over two-site 
(thick solid line) and one-site (thin solid line) channels, as a function 
of temperature. Dotted lines show $F(T)$ for direct tunneling 
and Ohmic conduction. Dashed line shows the averaged two-site current
$I(T)$ normalized by $I(0)$. 
} 
\label{F(T)}\end{figure}

        In the case when all $\Gamma_i^-$ are non-zero, it is more natural 
to use the general master-equation formalism for the average current 
\cite{Av-Likh} and spectral density.\cite{Kor-94} 
	We have developed a numerical code and integrated 
over different two-site channels 
in the same way as above, just using the numerical results for $I$ and 
$S_I(0)$ instead of Eqs.\ (\ref{cur}) and (\ref{S_I}). (One more difference
from the zero-temperature case is the separate integration 
over $\varepsilon_1$ and $\varepsilon_2$.) 
The dashed line in Fig.\ \ref{F(T)} shows 
the dependence of the ratio $g \equiv I_\Sigma (T)/I_\Sigma(0)$ 
on the normalized temperature $T/eV$. [We neglect the weak temperature 
dependence of $\tilde x_2$ and actually calculate the dependence of 
the numerical factor in Eq.\ (\ref{Isum2}).]  
	The asymptote 
at $T\gg eV$ is $g=21.7 \, (T/eV)^{4/3}$, so the conductance $G$ is equal 
to $G=113.6 \, e^2 \Gamma_0 \, n^2 A a^3 {\tilde x}_2 \, \exp (-2d/3a)\, 
\alpha^{1/3} T^{4/3}$, where $\tilde x_2$ can be approximated as
$(d/3)+(a/3)\ln (\alpha T)$ (the scaling $T^{4/3}$ is the same as in 
the model without correlations\cite{Glazman}).  

        The Fano factor averaged over different channels, as a function
of $T/eV$ is shown in Fig.\ \ref{F(T)} by the thick solid line. 
The low-temperature value is given by Eq.\ (\ref{Faver}), while the 
high-temperature asymptote, $\overline{F} = 2T/eV$ (lower dotted line) 
can be easily derived from the Nyquist formula. 

        It is interesting to compare the temperature dependence
of $\overline{F}$ for two-site and one-site channels. In the latter case
we still use Eqs.\ (\ref{Gamma_in}) and (\ref{Gamma_out}) for the tunneling 
rates similar to Ref.\ \cite{Nazarov}. The thin solid line in Fig.\ 
\ref{F(T)} shows the average Fano factor for one-site channels, as a function
of $T/eV$ (this curve in other coordinates has been calculated in Ref.\ 
\cite{Nazarov}).  The low temperature value is $\overline{F} =3/4$,  
while the high-temperature asymptote, $\overline{F} = 2T/eV$, 
is the same as for two-site channels and direct-tunneling case.
(The result for the direct tunneling \cite{}, $F=\coth (eV/2T)$, is 
shown for comparison by the upper dotted line.) Obviously, with the increase 
of the number $N$ of sites in the channel the average Fano factor 
decreases. However, its dependence on $N$ seems to saturate rapidly, 
as indicated by the small difference between the results for one-site and 
two-site channels. So, even for large $N$ one should expect  
the dependence $\overline{F}(T)$ to deviate significantly at $T\lesssim eV$
from the result for an Ohmic conductor, $F=2T/eV$. This can be explained 
by the fact that the 1D chains of sites with ``soft'' (not strong) bottlenecks 
still give considerable contribution to the total current, while the Fano
factor for such chains is comparable to unity. 
(The situation is different \cite{Sverdlov} for 2D or 3D hopping 
because the percolation cluster does not have bottlenecks at the size 
scale much larger than the correlation length of the cluster. 
As a consequence, for sufficiently large samples we expect $F=2T/eV$.)

\begin{figure} 
\begin{center}
\vskip-0.2cm
\epsfig{figure=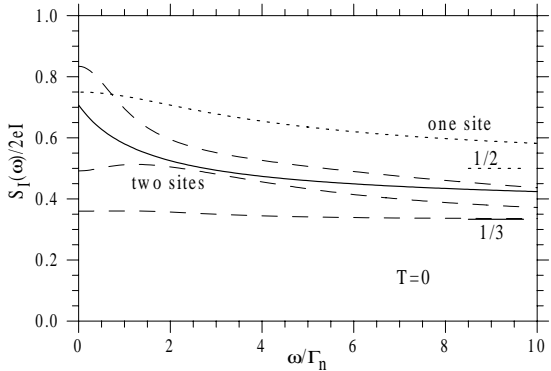,scale=0.4} 
\vskip-0.2cm
\end{center}
\caption{Solid line: the frequency dependence of the normalized current 
spectral density $S_I(\omega )/2eI$ averaged over two-site channels. 
Dashed lines correspond to particular channels (see text). 
The dotted line shows $S_I(\omega )/2eI$ averaged over one-site channels. 
 }
\label{S(w)}\end{figure}

        So far we have discussed only the current spectral density 
at zero frequency. Our computer code can also treat the finite-frequency
case. At finite frequency $\omega$ it is necessary to specify where 
the current 
is measured. We have considered the current in the electrodes and assumed 
natural 
electrostatics when the electron hop through $i$th gap transfers charge 
$q_i=ex_i/d$ in the electrodes. The spectral density at finite $\omega$ 
depends not only on tunneling rates $\Gamma_i^\pm$ but also on $q_i$ and thus 
on the positions of sites. For averaging over the two-site channels we have
used the approximation $q_i\simeq e{\tilde x}_i/d \simeq e/3 $.
	The solid line in Fig.\ \ref{S(w)} shows the frequency 
dependence $S_{I,\Sigma}(\omega )/2eI_\Sigma$ for $T=0$. 
The frequency scale is determined by tunneling rates, so for normalization
we have used $\Gamma_n \equiv \Gamma_0 \exp(-2d/3a) (\alpha eV)^{1/3}$. 
For comparison, we also show $S_I(\omega )/2eI$ for 
the uniform channel, $\Gamma_1=\Gamma_2=\Gamma_3$ (lower dashed 
line) and for two nonuniform channels:   
$\Gamma_1=\Gamma_2=0.1\Gamma_3$ (middle dashed line) and
$10\Gamma_1=\Gamma_2=0.1\Gamma_3$ (upper dashed line) 
at $\Delta \varepsilon =eV$. 
Notice that the solid line (averaged noise) has a finite slope 
at $\omega =0$ (even though 
the slope is zero for each individual channel) and approaches the 
high-$\omega$ asymptote of $1/3$ as $\omega^{-1/2}$. 
 The dotted line in Fig.\ \ref{S(w)} shows the ratio $S_I(\omega )/2eI$ 
averaged over one-site channels 
[then $\Gamma_n \equiv \Gamma_0\exp (-d/a)$], which can be calculated
analytically: $1/2+[(\omega^2\Gamma_n^{-2} +4)^{1/2}-2]\,\Gamma_n^2/\omega^2$.

        Finally, let us briefly consider the effect of 
electron spin using a simple model. Assuming the double degeneracy
due to spin (but still allowing at most one electron per site), 
we should double the tunneling rates $\Gamma_1^+$ 
and $\Gamma_3^-$ [see Eq.\ (\ref{Gamma_in})], while leaving all other rates 
unchanged. At zero temperature this will lead to a trivial extra 
factor $2^{1/3}$ in Eq.\ (\ref{Isum2}) and a very small change 
of ${\tilde x}_2$, while the average Fano factor given by Eq.\ 
(\ref{Faver}) does not change. The calculations 
at finite temperature show that $\overline{F}$   
is a little larger  
in the case of double degeneracy compared to the spinless case,
however, the difference is so small 
that the corresponding curves in Fig.\ \ref{F(T)} cannot be 
distinguished. 
The maximum difference $\Delta {\overline F} \approx 5\times 10^{-4}$ is 
achieved at $T/eV\approx 0.3$, 
while at $T\gg eV$ the difference approaches zero.
A similar very weak dependence on the spin degeneracy for 
one-site channels has been reported in Ref.\ \cite{Nazarov} 
(we have found the maximum difference $\Delta \overline F=1.12\times 10^{-3}$ 
at $T/eV \approx 0.33$).  

	In conclusion, we have studied the shot noise of two-site 
hopping channels. The different average Fano factor 
and different frequency dependence of the noise 
in comparison with one-site channels and direct tunneling 
can in principle be verified experimentally (using  
the difference of the temperature dependence of the average current).

Fruitful discussions with K. K. Likharev, V. A. Sverdlov,
V. V. Kuznetsov, and K. A. Matveev  are 
gratefully acknowledged. 
This work was supported in part by the Engineering Research Program of 
the Office of Basic Energy Sciences at the Department of Energy.


\begin{references}
\bibitem{interaction} To demonstrate the convenience of the method, 
        let us extend the model taking into account the interaction  
        of electrons on two sites. Similar to Ref.\ 
        \protect\cite{Matveev} let us replace the rate for 
        transition $01 \rightarrow 11$ with ${\tilde \Gamma}_1$ 
        and the rate for transition $11\rightarrow 10$ with 
        ${\tilde \Gamma}_3$. We still have two types of blocks,
        and calculating ${\overline \tau}$ we obtain the 
        average current,   
        $I = e({\tilde\Gamma}_1 +\Gamma_3)/\left[ 1+
        {\tilde\Gamma}_1 ({\tilde\Gamma}_3^{-1} + \Gamma_2^{-1} ) + 
        \Gamma_3 (\Gamma_1^{-1} + \Gamma_2^{-1} ) \right]  $,
        in a much simpler way than solving the master 
        equation conventionally.\protect\cite{Av-Likh} The expression 
        for $S_I(0)$ can be also easily obtained.
\bibitem{e-e_2} The equations can be easily modified to take into account
	the electron-electron interaction in the way similar to the remark 
	\protect\cite{interaction} above.
\end{references}
\end{document}